\documentclass[11pt]{article}
\pdfoutput=1
\usepackage{geometry}                
\usepackage{graphicx}
\usepackage{amssymb}
\usepackage{amsmath}
\usepackage{epstopdf}
\usepackage{amscd}
\usepackage{mathtools}
\usepackage{slashed}
\usepackage{bbm}
\usepackage{upgreek}
\usepackage[utf8]{inputenc}
\usepackage[matrix,arrow]{xy}
\usepackage{arydshln}
\usepackage[utf8]{inputenc}

\usepackage{stolenstyle}

\usepackage[linktocpage=true]{hyperref}
\hypersetup{
colorlinks=true,
citecolor=blue,
linkcolor=blue,
urlcolor=blue}

\DeclareFontFamily{OT1}{pzc}{}
\DeclareFontShape{OT1}{pzc}{m}{it}{<-> s * [1.10] pzcmi7t}{}
\DeclareMathAlphabet{\mathpzc}{OT1}{pzc}{m}{it}

\newcommand\T{\rule{0pt}{2.6ex}}       
\newcommand\B{\rule[-1.2ex]{0pt}{0pt}} 

\newcommand{\overbar}[1]{\mkern 1.5mu\overline{\mkern-1.5mu#1\mkern-1.5mu}\mkern 1.5mu}
\def\Gbar{\overbar{G}}

\def\Dbar{\overbar{D}}

\def\ebar{\overbar{e}}
\def\zbar{\overbar{z}}

\def\kappabar{\overbar{\kappa}}

\def\epsbar{\overbar{\epsilon}}

\def\ie{{\it i.e.}}
\def\eg{{\it e.g.}}

\def\tr{ \, \textrm{tr} \, }

\def\Tr{ \, \textrm{Tr} \, }
\def\STr{ \, \textrm{STr} \,}

\def\diag{ \, \textrm{diag}}

\def\NN{\mathcal{N}}

\def\half{\frac{1}{2}}

\def\KK{\mathcal{K}}

\def\RR{\mathcal{R}}

\def\VV{\mathcal{V}}

\def\pd{\partial}

\def\preprint{}

\title{Nonabelian Probes In Holography }
\author[a] {Sophia K.~Domokos}
\author[b] {and Andrew B.~Royston}

\abstract{We find the range of parameters for which the open string physics on probe D$q$-branes in the near-horizon geometry of D$p$-branes decouples from gravity, and is well-approximated by a $(q+1)$-dimensional supersymmetric Yang--Mills--Higgs theory on a rigid curved spacetime.  We study the vacua of these theories, which include moduli spaces of instantons, monopoles, and vortices.  This intricate structure is made possible through couplings to the background Ramond-Ramond flux.  The probe brane theories we study provide holographic descriptions of defects in dual field theories.} 

\affiliation[a]{Department of Physics, New York Institute of Technology, 16 W. 61st Street, New York, NY 10023}
\affiliation[b]{ Department of Physics, Penn State Fayette, The Eberly Campus, 2201 University Drive, Lemont Furnace, PA 15456 }
\emailAdd{sdomokos@nyit.edu}
\emailAdd{abr84@psu.edu}

\begin{document}

\maketitle
\parskip 7pt

\section{Introduction and Summary of Results}

Supersymmetric Yang--Mills--Higgs theories on holographic backgrounds have applications in both mathematics and physics.

On the mathematics side, BPS equations in Yang--Mills--Higgs theories take the form of generalized self-duality equations.  These are first order PDE's that generalize the instanton equation and have seen wide-ranging applications, like the four-manifold invariants of \cite{MR1066174}, the geometric Langlands program of \cite{Kapustin:2006pk}, and the gauge theoretic construction of Khovanov homology \cite{Witten:2011zz}, to name a few.  One often studies these equations on a manifold with boundary -- a set-up that arises naturally in the holographic context.  Holography thus provides a new tool for exploring the consistency of boundary value formulations.

On the physics side, Yang--Mills--Higgs theories on holographic backgrounds can yield insights into real-world systems.  The holographic QCD program, for instance, has generated a workable framework for studying glueballs and mesons, but baryons remain more elusive.  While glueballs and mesons are dual, respectively, to perturbative closed string states and perturbative open string states on probe D-branes, baryons are realized as solitons of the open string fields.  Progress in this area is rendered more difficult by the complicated supergravity backgrounds in which the probe branes, in top-down models like \cite{Sakai:2004cn,Witten:1998zw}, are embedded. Most precise calculations in this area rely on numerics \cite{Cherman:2011ve, Bolognesi:2013nja} and approximations \cite{Sakai:2004cn,Hong:2007kx}.  (The various approaches are nicely summarized in \cite{Bolognesi:2013nja}.)  {\em Supersymmetric} gauge theories on holographic spacetimes offer an alternative approach: in these controlled settings, one might hope to extract exact results about the structure (and even the dynamics \cite{Manton:1981mp}) of solitons, which may then shed light on holographic baryons as well.

Realizing supersymmetric gauge theories on rigid curved backgrounds is a nontrivial task.
 The authors of \cite{Festuccia:2011ws,Dumitrescu:2012ha} initiated a program for constructing such theories that involves a limit of off-shell supergravity backgrounds.  This approach and its generalizations -- which subsume the classic notion of topological twisting -- have successfully constructed many supersymmetric theories in $D \leq 4$, including theories with eight supercharges \cite{Butter:2015tra}.  The extent to which these theories overlap with the class of supersymmetric theories arising from D-branes in string theory flux backgrounds is unclear \cite{Maxfield:2015evr,Triendl:2015lka}.
  
 Fortunately, nonabelian D-brane actions \cite{Myers:1999ps} and $\kappa$-symmetry in curved backgrounds \cite{Aganagic:1996pe,Aganagic:1996nn, Cederwall:1996ri, Bergshoeff:1996tu} provide the tools needed to construct such actions directly.   We did this in \cite{Domokos:2017wlb} for the case of D5-branes probing the near-horizon geometry of D3-branes.  The abelian version of this system, first studied by \cite{Karch:2001cw,DeWolfe:2001pq, Erdmenger:2002ex}, is holographically dual to a defect CFT. In  \cite{Domokos:2017wlb}, we also found that the nonabelian probe brane theory has a highly nontrivial space of vacua, as well as systems of BPS equations  for finite-energy soliton configurations. The coupling of the probe fields to the background flux via the Chern-Simons part of the D-brane action plays an essential role in producing this structure.

In this paper we extend aspects of this analysis to supersymmetric D$p$/D$q$ intersections.  Rather than constructing complete actions (which could straightforwardly be done using the techniques of \cite{Domokos:2017wlb}), we focus on two key results:
\begin{enumerate}
\item In a broad class of supersymmetric D$p$/D$q$ intersections we find a regime of parameter space in which the low-energy effective action for open string modes on the probe D$q$-branes is well-approximated by a supersymmetric Yang--Mills--Higgs theory on a rigid, curved background.  We can decouple the closed string modes and suppress higher order derivatives in open string modes by appropriately tuning the number of D$p$-branes ($N_c$), the number of D$q$-branes ($N_f$), and the effective D$p$-brane 't Hooft coupling.

We find that the probe limit for these nonabelian probes -- in which one can safely neglect backreaction --  is in general modified from the oft-quoted $N_f\ll N_c$ by factors of the 't Hooft coupling. 

\item We uncover intricate spaces of zero-energy vacua in these probe D$q$-brane theories.  For D$p$/D$(p+4)$, D$p$/D$(p+2)$, and D$p$/D$p$ intersections, these include moduli spaces of instantons, monopoles, and vortices, respectively.  As in the D3/D5 case of \cite{Domokos:2017wlb}, the existence of such vacua is made possible through couplings between the background fluxes and the probe fields via the Chern-Simons action. Specifically, the terms that come from the Chern-Simons action provide cross terms that, together with the standard kinetic terms from the DBI action, allow one to write the Hamiltonian as a sum of non-trivial squares. Setting these squares to zero yields a set of Bogomolny-type equations, that are translationally invariant along the directions common to both stacks of branes.
\end{enumerate}

The paper is organized as follows: In Section \ref{sec:YM} we determine the regime of parameters that isolates classical Yang--Mills--Higgs theory on the probes; in Section \ref{sec:vacua} we elucidate vacuum structures associated with near-horizon limits of D$p$/D$(p+4)$, D$p$/D$(p+2)$, and D$p$/D$p$ intersections; in Section \ref{sec:concl} we conclude and discuss directions for future work.

%
\section{Isolating Yang--Mills--Higgs Theory on Probe Branes}\label{sec:YM}

\subsection{Notation and Setup}

We consider orthogonal intersections of $N_c$ ``color'' D$p$-branes with $N_f$ ``flavor" D$q$-branes.  The D$p$-branes generate a supergravity background with nontrivial metric.  Tildes indicate coordinates of the full (asymptotically flat) spacetime. Coordinates without tildes parameterize the near-horizon region.

Coordinate assignments are summarized in Table \ref{notationtable}.  The D$q$-branes span the $x^0,x^1\dots, x^q$ directions, while the D$p$-branes span $x^0,x^1,\ldots, x^d, x^{9-(p-d)+1},\ldots,x^{9}$.  Axes parallel to both stacks of branes are labelled $x^\mu$, where $\mu = 0, 1, \dots d$.  Axes parallel to the D$q$-branes and perpendicular to the D$p$-branes are labeled $\vec{r} = (r_1,\ldots,r_{q-d})$, with $r_i = x^{d+i}$ and with $r\equiv \sqrt{\vec{r}\cdot\vec{r}}$.   Axes transverse to both stacks are labelled $\vec{z}=(z_1, \dots, z_{9-(p+q-d)})$, with $z_i = x^{q+i}$ and with $z \equiv \sqrt{\vec{z} \cdot \vec{z}}$.  Finally, axes parallel to the D$p$-branes and perpendicular to the D$q$-branes are labeled $\vec{y}=(y_1, \dots, y_{p-d})$, with $y_i = x^{9-(p-d)+i}$.

 We will also use $x^a$ to refer to all of the directions parallel to the D$q$-branes, and $x^m$ to refer to all of the directions transverse to them.

The D$p$/D$q$ intersections we consider are $1/4$-BPS, preserving eight supercharges. This imposes the requirement that $(p-d)+(q-d)$ be divisible by four \cite{Polchinski:1998rr}.  In what follows, we allow the stacks to be separated by a constant displacement, $\vec{z}_0$, along the directions transverse to both. This does not break any additional supersymmetries, and should correspond to a relevant deformation of the dual defect field theory (see, \eg\ \cite{Karch:2002sh,Yamaguchi:2002pa}). Our analysis also applies to ``transversal'' intersections having $p+q-d = 9$, where there are no directions along which to separate the stacks.

\begin{center}
\begin{table}
\begin{tabular}{c | c c c c ; {2pt/2pt} c c c ;{2pt/2pt} c c c ;{2pt/2pt} c c c}
  & \small{0} & \small{1} & \small{\dots} & \small{$d$} & \small{$d+1$} & \small{\dots} & \small{$q$} & \small{$q+1$} & \small{\dots} & \small{$9-(p-d)$} & \small{$10-(p-d)$} & \small{\dots} & \small{9}  \T\B\B \\
  \hline
 D$p$ & X & X & \dots & X &  & &  & & & & X & \ldots & X   \T\T\B  \\
 D$q$ & X & X & \dots & X & X & \ldots & X & & &  & & &  \T\B\B \\ \hline
      & \multicolumn{4}{c ;{2pt/2pt}}{$x^\mu = (t, x_i)$} & \multicolumn{3}{c}{$\vec{r} = (r_i)$} & \multicolumn{3}{;{2pt/2pt} c ;{2pt/2pt}}{$\vec{z}=(z_i)$} &  \multicolumn{3}{;{2pt/2pt} c }{$\vec{y}=(y_i)$}  \T\T\B \\ \hline 
      & \multicolumn{7}{c ;{2pt/2pt}}{$x^a$} & \multicolumn{6}{c }{$x^m$} 
       \T\T
 \end{tabular}
 \caption{D$p$/D$q$ intersection}\label{notationtable}
 \end{table}
\end{center}

\subsection{Background Geometry and Near-Horizon Limit}

We now review the D$p$-brane background and its near-horizon limit. In this subsection we focus exclusively on the D$p$-branes and ignore the D$q$-branes.

The extremal $N_c$ D$p$-brane supergravity solution includes a nontrivial background metric, dilaton, and Ramond-Ramond $(p+1)$-form potential.  In string frame, the solution for $p<7$ is given by  \cite{Horowitz:1991cd, Duff:1993ye}:
\begin{align}\label{eq:FullD$p$Bckgd}
ds^2 &= f_p^{-1/2}\left( dx^\mu dx^\nu \eta_{\mu\nu} + d\vec{{y}} \cdot  d\vec{{y}} \right)+f_p^{1/2}\left(d\vec{\tilde{r}}\cdot d\vec{\tilde{r}} + d\vec{\tilde{z}}\cdot d\vec{\tilde{z}} \right)~, \cr
e^{-2(\phi-\phi_\infty)}&=f_p^{(p-3)/2} ~, \cr
C_{0\cdots p} &= - e^{-\phi_\infty}\left( f_p^{-1}-1\right)~,
\end{align}
where
\begin{equation}\label{fp}
f_p = 1+\left(\frac{L_p}{\tilde{v}}\right)^{p+1}~,
\end{equation}
with $\tilde{v} \equiv \sqrt{ \tilde{r}^2+\tilde{z}^2}$ the radial direction transverse to the D$p$-branes, and $e^{\phi_{\infty}} = g_s$ the asymptotic value of the dilaton. The length scale $L_p$ is given by 
\begin{equation}
L_p^{7-p}=(4\pi)^{\frac{5-p}{2}}\Gamma\left( \frac{7-p}{2}\right) N_cg_s\ell_s^{7-p}~.
\end{equation}

We now take the standard AdS/CFT low energy limit \cite{Itzhaki:1998dd, Boonstra:1998mp, Peet:1998wn}. From the closed string point of view, this means zooming in on the near-horizon geometry of the stack. From the open string point of view, it reduces the low-energy effective theory on the D$p$-branes to maximally supersymmetric $(p+1)$-dimensional Yang--Mills--Higgs.  

We define the limit in terms of a scale $\mu$ that acts like a renormalization scale for the field theory on the D$p$-branes, so that the low-energy limit is given by $\mu\ell_s\rightarrow 0$, with the dimensionless D$p$-brane Yang--Mills coupling,
\begin{equation}
g_{p}^2 :=(2\pi)^{(p-2)} g_s (\mu\ell_s)^{p-3} ~,
\end{equation}
held fixed.  Note that this means that $g_s$ must go to zero or blow up for $p \neq 3$. (The dimensionful D$p$-brane Yang--Mills coupling used in \cite{Itzhaki:1998dd} and elesewhere is equal to $g_{p}^2 \mu^{-(p-3)}$.)

We also need to keep the energy scale of characteristic processes in the field theory fixed relative to $\mu$.  There are two natural energy scales to consider. One energy scale, denoted $U$ below, is associated with the Higgs vev in the field theory, and is equivalent to the energy of a string stretched between the D$p$ stack and an additional probe D$p$-brane, separated by a distance $\tilde{v}$ \cite{Itzhaki:1998dd}.  The other, denoted $E$ below, is a cutoff energy in the field theory imposed, for instance, to regularize the computation of the density of states.  As explained by \cite{Susskind:1998dq}, cutting off the field theory in this way is dual to cutting off the AdS space at the radius $\tilde{v} = \tilde{v}(E)$.  The scale $E$ also determines the characteristic radial dependence of supergravity modes \cite{Peet:1998wn}. These two energy scales are given by

\begin{equation}\label{Ewithvtilde}
U = \frac{\tilde{v}}{\ell_s^2}  \qquad \textrm{and} \qquad  E = \frac{(\tilde{v}/\ell_s^2)^{\frac{5-p}{2}}}{\sqrt{N_c g_{p}^2\mu^{-(p-3)}}} =\frac{\mu}{{\sqrt{N_c g_{p}^2}}} \left( \frac{\tilde{v}}{\mu\ell_s^2}\right)^{\frac{5-p}{2}}~.
\end{equation}

When $p < 5$, both of these scales are held fixed by holding $\tilde{v}/(\mu \ell_s^2)$ fixed. This implies, in particular, that $\tilde{v}/\ell_s \to 0$.  For $p=5$, the supergravity energy-distance relation degenerates. While there is a holographic dual in terms of little string theory \cite{Aharony:1998ub}, it is not a local quantum field theory.  For $p=6$, these two probes of the field theory energy have opposite behavior with respect to the radial coordinate $\tilde{v}$, and there is not expected to be a field theory dual \cite{Itzhaki:1998dd,Peet:1998wn}.  This is the case in general for $p\geq 6$.  From now on, we restrict our attention to $p \leq 4$.

From the perspective of the geometry, the limits $\mu \ell_s \to 0$ and $\tilde{v}/\ell_s \to 0$ with $g_{p}^2$ and $E/\mu$ fixed constitute the classic near-horizon limit ($L_p/\tilde{v} \to \infty$).  In order to facilitate taking this limit on the supergravity side, we introduce a new radial coordinate that remains finite:
\begin{equation}\label{vdef}
\mu^2 v =\frac{E}{\alpha_p\sqrt{d_p}} ~,
\end{equation}
where
\begin{equation}\label{eq:alphapD$p$}
\alpha_p :=  \frac{2}{5-p}\qquad \text{and}\qquad d_p := 2^{7-2p}\pi^{\frac{9-3p}{2}}\Gamma\left( \frac{7-p}{2}\right)~
\end{equation}
are included for later convenience.  In terms of the Cartesian coordinates $\vec{r}$ and $\vec{z}$ that parameterize directions transverse to the D$p$-branes, equation \eqref{vdef} originates from a uniform rescaling
\begin{equation}
r_i = \frac{1}{\sqrt{\alpha_{p}^2 d_p N_c g_{p}^2}} \left( \frac{\tilde{v}}{\mu \ell_{s}^2} \right)^{\frac{3-p}{2}} \frac{\tilde{r}_i}{(\mu \ell_{s})^2}~, \qquad  z_i = \frac{1}{\sqrt{\alpha_{p}^2 d_p N_c g_{p}^2}} \left( \frac{\tilde{v}}{\mu \ell_{s}^2} \right)^{\frac{3-p}{2}} \frac{\tilde{z}_i}{(\mu \ell_{s})^2} ~,
\end{equation}
with $v = \sqrt{ r^2 + z^2}$.

In terms of these new coordinates and the dimensionless coupling $g_p$, the near-horizon $N_c$ D$p$-brane background is
\begin{align}\label{NearHorizon}
ds^2 &= G_{MN}dx^M dx^N \cr
&\to \lambda_G (\mu v)^{\frac{p-3}{5-p}}\left\{ (\mu v)^2\left( \eta_{\mu\nu}dx^\mu dx^\nu + d\vec{y}\cdot d\vec{y} \right)+\frac{dv^2}{(\mu v)^2} + \frac{1}{\alpha_p^2\mu^2}d\Omega^2_{8-p}\right\} \cr
&=:  \lambda_G \, \Gbar_{MN}dx^M dx^N \cr
e^\phi &\to \lambda_\phi (\mu v)^{\frac{(p-3)(7-p)}{2(5-p)}} \cr
C_{0\cdots p}&\to - \lambda_C (\mu v)^{\frac{(p-3)^2}{5-p}+(p+1)} ~,
\end{align}
where
\begin{align}
\lambda_G &:= (\alpha_p^{7-p}d_p)^{\frac{1}{5-p}}(N_c g_p^2)^{\frac{1}{5-p}} (\mu\ell_s)^2 ~, \\
\lambda_\phi &:= \frac{\left( \alpha_p^{7-p}d_p\right)^{\frac{p-3}{2(5-p)}}}{(2\pi)^{p-2}} \frac{1}{N_c} \left( N_c g_p^2 \right)^{\frac{7-p}{2(5-p)}}~, \label{eq:dilaton}\\
\lambda_C &:= (2\pi)^{p-2} \left( \alpha_p^{7-p}d_p\right)^{\frac{2}{5-p}} (\mu\ell_s)^{p+1} N_c\left( N_c g_p^2 \right)^{\frac{p-3}{5-p}} ~.
\end{align}
Note that the near-horizon dilaton is finite in the limit and the Ramond-Ramond $p$-form goes as $(\mu \ell_s)^{p+1}$. The near-horizon metric goes like $(\mu \ell_s)^2$, and is conformal to $AdS_{p+2}\times S^{8-p}$ with radii $\mu^{-1}$ and $(\alpha_p\mu)^{-1}$, respectively.  The fact that scale transformations $v \mapsto \lambda v$ are not isometries when $p \neq 3$ corresponds to the fact that dual QFT is not a CFT.

Let us briefly review the regime of validity of these supergravity solutions -- that is, the range of $v$ for which both the dilaton and the Ricci scalar are small \cite{Itzhaki:1998dd}. The dilaton is given in \eqref{eq:dilaton} while the Ricci scalar measured in string units goes as
\begin{equation}
\ell_s^2 R \sim \left( N_cg_p^2\right)^{-\frac{1}{5-p}} (\mu v)^{-\frac{p-3}{5-p}}~.
\end{equation}
Both quantities are small for the following ranges of the radial coordinate $v$ in terms of $N_c$ and $g_p$:
\begin{align}\label{sugraconds}
&\text{for}~~ p=4:\qquad\frac{1}{N_cg_4^2}\ll \mu v \ll \frac{N_c^{2/3}}{N_cg_4^2} \cr
&\text{for}~~ p=3:\qquad 1 \ll N_cg_3^2 \ll N_c~, \qquad (\textrm{no restriction on $v$})~, \cr
&\text{for}~~ p<3:\qquad\frac{(N_cg_p^2)^{\frac{1}{3-p}}}{N_c^{\frac{2(5-p)}{(3-p)(7-p)}}} \ll \mu v \ll (N_c g_p^2)^{\frac{1}{3-p}}  ~.
\end{align}
For $p=4$ the condition on the dilaton leads to the upper bound and the condition on the curvature to the lower bound.  For $p<3$ these conditions reverse roles.  When the dilaton becomes large it is possible to switch to a dual description involving a different supergravity background.  When the curvature becomes large, meanwhile, the dual field theory description becomes weakly coupled \cite{Itzhaki:1998dd}.   Note that the $p=3$ conditions do not constrain $v$ at all, giving instead conditions on the 't Hooft coupling and $N_c$. While here we focus on the regime where the above solutions are appropriate, we expect that certain quantities, like BPS field configurations on probe branes in these geometries, will extend beyond the strict limits above. 

Having reviewed the low-energy limit, we now determine the effective Newton constant $\kappabar$, which dictates the coupling between the bulk supergravity (closed string) fluctuations and the open string fluctuations on the probe branes. We expand the Type II supergravity action around the background in canonically normalized metric fluctuations $h_{MN}$, so that
\begin{equation}
G_{MN} = G^{background}_{MN}+ \kappabar  e^{\phi/2}h_{MN},
\end{equation}
where the factor of $e^{\phi/2}$ takes us from string frame to Einstein frame. The constant factors in the background metric and dilaton combine with the usual 10d  Newton constant $\kappa^2_{10}=\frac{1}{2}(2\pi)^7 \ell_s^8$ to give
\begin{equation}\label{eq:kappabar}
\kappabar = \frac{\kappa_{10}\lambda_\phi}{\lambda_G^2} \sim \mu^{-4}\frac{1}{N_c} (N_cg_p^2)^{-\frac{p-3}{5-p}}~.
\end{equation}
In the last step we neglected numerical factors irrelevant to our analysis below. Note that the factors of $\ell_s$ cancel, as they should.

\subsection{Adding Probe Branes}
Having established the background, we now add $N_f$ probe D$q$-branes on which we will isolate the supersymmetric Yang--Mills--Higgs theory of interest.

The parametrization of this orthogonal intersection is given in Table \ref{notationtable}.  These probes break half of the supersymmetry of the D$p$-brane background.  As mentioned above, this restricts the D$p$/D$q$ intersections we can consider to those with 0, 4, or 8 ND directions (\ie\ $(p-d)+(q-d)$ divisible by four) \cite{Polchinski:1998rr}. We further restrict to intersections in which the D$q$-branes have at least one direction transverse to the color D$p$'s, so that they extend to the holographic boundary. 

The bosonic massless open string degrees of freedom on the D$q$-branes consist of a nonabelian $\mathrm{U}(N_f)$ gauge field $A_a$ and adjoint-valued transverse scalars $X^m$.  We are interested in a regime of parameter space where (1) the couplings of bulk supergravity fluctuations to the probe brane theory are suppressed, and (2) the action on the probes is well-approximated by a (classical) Yang--Mills--Higgs theory.  

The low-energy effective action of the bosonic massless open string modes is captured by the nonabelian D-brane action of Myers \cite{Myers:1999ps}.\footnote{The Myers action is known to give results incompatible with the low-energy limit of string scattering amplitudes beyond certain orders in both the fieldstrength and derivative expansions; these issues will not be relevant here.  See \cite{Domokos:2017wlb} for a fuller discussion and references.}  

As we will demonstrate now, the expansion of the Myers action in open string modes is governed by two parameters, $g_{q(p)}$ and $\epsilon_p$. Here, $g_{q(p)}$ is the effective Yang--Mills coupling on the probe D$q$-branes (the $(p)$ subscript is intended to indicate its dependence on the D$p$-brane background), while $\epsilon_p$ governs the $\alpha' = \ell_{s}^2$ expansion.\footnote{$\epsilon_p$ was denoted $\epsilon_{\rm op}$ in \cite{Domokos:2017wlb}, where we focused exclusively on the D3/D5 intersection.}  The expansion in closed string fluctuations is governed by $\kappabar$ \eqref{eq:kappabar}.

The Myers action is a sum of Dirac-Born-Infeld (DBI) and Chern-Simons (CS) actions:
\begin{align}
S^{\rm bos}_{{\rm D}q} =& S_{\rm DBI} + S_{\rm CS}~, \qquad {\rm with} \cr
S_{\rm DBI} =&~ T_{{\rm D}q} \int d^{q+1}x  \STr e^{-\phi} \sqrt{-\det \left( P[G_{ab}]-i (2\pi \alpha')F_{ab} \right) \det\left(\delta^{m}_{\phantom{n}n} -i  (2\pi\alpha)^{-1} [X^m, X_n]\right) }~. \cr \label{DBIfull}
\end{align}
For the purposes of this section, we will focus on the DBI action. We will return to the CS action in the next section.

The brane tension is $T_{{\rm D}q} = 2\pi/(2\pi\ell_s)^{q+1}$, and the quantity $P[T_{MN...Q}]$  denotes the gauge-covariant pullback $P$ of a bulk tensor $T_{MN...Q}$ to the worldvolume of the D$q$-branes. For instance, the pullback of the bulk metric $G_{ab}$ is
\begin{equation}\label{PofG}
P[G_{ab}] = G_{ab} -i (D_a X^m) G_{mb} -i G_{am} (D_b X^m)  - (D_a X^m) G_{mn} (D_b X^n) ~,
\end{equation}
with $D_a = \pd_a + [A_a,\,\cdot\,]$\footnote{We use conventions in which the adjoint-valued fields are represented by anti-hermitian matrices, and $\Tr\{~,~ \}:=-\tr_{N_f}\{~,~ \}$ is a positive-definite Killing form on the Lie algebra.}.  The closed string fields are to be taken as functionals of the matrix-valued coordinates, $G_{MN}(x^P) \to G_{MN}(x^a;-iX^m)$, defined by power series expansion.  The `STr' stands for a fully symmetrized trace, defined as in \cite{Myers:1999ps}. For the terms we will consider below, it reduces to an ordinary trace.  

Our first goal is to isolate the kinetic terms for $A_a$ and $X^m$ in order to determine the effective Yang--Mills coupling $g_{q(p)}$ and the appropriate normalization of the $X^m$. To that end,
we expand the determinants in the DBI action:
\begin{align}\label{DBI}
S_{\rm DBI} \supset &~ -T_{{\rm D}q} \int d^{q+1} x e^{-\phi} \sqrt{-\det(G_{ab})} \times \cr
&\qquad\qquad\qquad\times \Tr \bigg\{ \frac{(2\pi \alpha')^2}{4} G^{ab} G^{cd} F_{ac} F_{bd}  
+ \half G^{ab} G_{mn} D_{a} X^m D_{b} X^n  \bigg\}~. \qquad
\end{align}
Here we evaluate the pulled-back metric and dilaton at $\vec{z}=\vec{z}_0$. 

Next we insert the near-horizon geometry \eqref{NearHorizon} for the closed string background. It is useful to define the induced rescaled metric on the worldvolume $g_{ab} := P[\Gbar (x^a, -iX^m=x_0^m)]_{ab}$, such that
\begin{equation}
g_{ab} dx^a dx^b = (\mu v|_{z_0})^{\frac{p-3}{5-p}} \left\{ (\mu v|_{z_0})^2  \eta^{\mu\nu} dx^\mu dx^\nu + \frac{1}{(\mu v |_{z_0})^2} \left[\frac{r^2 + \alpha_{p}^{-2} z_{0}^2}{r^2 + z_{0}^2}dr^2 + \frac{r^2}{\alpha_{p}^2} d\Omega_{q-d-1}^2 \right] \right\}~,
\end{equation}
where $v |_{z_0} \equiv \sqrt{r^2 + z_{0}^2}$.  This metric will be used to raise the $a,b$-type indices.  When $z_0 = 0$ this space is conformal to $AdS_{d+2} \times S^{q-d-1}$ with radii $\mu^{-1}$ and $(\alpha_p \mu)^{-1}$ respectively.  Nonzero $z_0$ gives a smooth deformation with the same asymptotics.    

We identify the Yang--Mills coupling on the probe branes in the near-horizon geometry using the coefficient of the gauge-field kinetic terms:
\begin{equation}\label{gqpdef}
g_{q(p)} := T_{{\rm D}q}^{-1/2} \lambda_{G}^{-\frac{(q-3)}{4}}\lambda_{\phi}^{1/2} (2\pi \alpha')^{-1}  \sim \frac{\mu^{\frac{3-q}{2}}}{\sqrt{N_c}} (N_c g_p^2)^{\frac{10-(p+q)}{4(5-p)}}~.
\end{equation}
Notice that the factors of $\ell_s$ have dropped out, illustrating that this coupling is finite in the low-energy/near-horizon limit.

Next, in order to ensure that the vevs of the scalar fields remain finite in this limit,
we introduce mass-dimension one scalars
\begin{equation}
\Phi^m := \frac{\lambda_G}{(2\pi \alpha')} X^m~.
\end{equation}
This proportionality factor also allows us to interpret $D_a \Phi_m$ as the off-diagonal components of a ten-dimensional fieldstrength, $F_{MN}$.

Armed with these definitions we return to \eqref{DBIfull} and carry out an $\alpha'$ expansion suited to the near-horizon geometry. Replacing the closed string fields with their near-horizon limits and the $X^m$'s with $\Phi^m$'s, we find that the $\alpha'$ expansion is an expansion in the open string variables $\{ F_{ab}, D_a\Phi^m, [\Phi^m,\Phi^n], \mu\Phi^m\}$, governed by the parameter 
\begin{equation}\label{epsp}
\epsilon_p := \frac{2\pi \alpha'}{\lambda_G} \sim (N_c g_{p}^2)^{-\frac{1}{(5-p)}} \mu^{-2}~.
\end{equation}
Including closed string fluctuations as well, the expansion of the bosonic Myers action takes the form
\begin{equation}\label{DBIexp}
S_{{\rm D}q}^{\rm bos} = - \frac{1}{\epsilon_{p}^2 g_{q(p)}^2} \int d^{q+1} x \sqrt{-\det(g_{ab})} (\mu v |_{z_0})^{\frac{(p-3)(p-7)}{2(5-p)}} \sum_{n_0,n_c=0}^{\infty} \epsilon_p^{n_o} \kappabar^{n_c} V_{n_o,n_c}~.
\end{equation}
Here $V_{n_o,n_c}$ is a sum of monomials with each term containing $n_o$ open string variables from the set $\{ F_{ab}, D_a\Phi^m, [\Phi^m,\Phi^n], \mu\Phi^m\}$, and $n_c$ closed string fluctuations  \cite{Domokos:2017wlb}.  The $\mu \Phi^m$ originates from the fact that the closed string fields are to be viewed as functionals of the transverse scalars when pulled back to the worldvolume, which must also be expanded. The overall factors of $\mu v$ arise from the background dilaton.
 
 The $V_{2,0}$ term comprises the Yang--Mills--Higgs theory we wish to isolate, and has the form
 \begin{align}\label{V20}
 V_{2,0} =& \Tr \bigg\{ \frac{1}{4} F_{ab}F^{ab} +\frac{1}{2}  \Gbar_{mn} D_{a} \Phi^m D^a \Phi^n + \frac{1}{2}M_{z_i z_j}(r) \Phi^{z_i}\Phi^{z_j}\cr 
 &\qquad+ \frac{1}{4}  \Gbar_{mn} \Gbar_{m'n'} [\Phi^{m}, \Phi^{m'}] [\Phi^n, \Phi^{n'}] 
\bigg\}+ V_{2,0}^{{\rm CS}}~.
 \end{align}

We defer discussion of the CS contribution to $V_{2,0}$ to the next section. 

The mass term for the $\Phi^{z_i}$ scalars vanishes for the D3/D5 system but is in general present. The important fact for us is that the $\Phi^{y_i}$ masses {\em always} vanish, as the background geometry depends on the $z_i$ but not on the $y_i$.

In order for the Yang--Mills--Higgs theory to provide a good leading order description of the physics, we need both higher order terms in the $n_o, n_c$ expansion, and the effects of lower order terms, to be suppressed. 

The higher order terms will be suppressed provided the expansion parameters at the scale of the field variations, $\mu^\prime$,  are small:
\begin{equation}
\epsilon_{p} \mu^{\prime 2} \ll 1~, \qquad g_{q(p)} \mu^{\prime (q-3)/2}  \ll 1~, \qquad\kappabar \mu^{\prime 4} \ll 1~.
\end{equation}
 In the following we will assume that $\mu'$ is parametrically the same as $\mu$, but the discussion can be modified accordingly if this is not the case.

Now consider the lower order terms. 
The $V_{0,0}$ term corresponds to the energy density of the background D$q$-branes, and plays no role.  The $V_{1,0}$ term vanishes because the supersymmetric D$q$-brane embedding on the fixed background satisfies the open string equations of motion.   $V_{1,1}$ terms are present, but only couple to the $\mathfrak{u}(1) \subset \mathfrak{u}(N_f)$ degrees of freedom due to the trace.  We focus here on the $\mathfrak{su}(N_f)$ sector, which decouples from the $\mathfrak{u}(1)$ sector at tree level.

Finally, there are the closed string tadpoles in $V_{0,1}$. These are present because we treat the D$q$-branes as probes and do not solve the full supergravity equations of motion.  The strength of these tadpoles is $N_f g_{q(p)}^{-2}\kappabar$, which can be large. However, their effects can only be transmitted to the $\mathfrak{su}(N_f)$ sector through a $V_{2,1}$ vertex, which goes as $g_{q(p)}^{-2}\kappabar$. Hence taking into account the canonical normalization of open string modes, the correction to \eg \  the gauge field propagator  goes as $N_f g_{q(p)}^2 (g_{q(p)}^{-2} \epsilon_{p}^{-2} \kappabar)(g_{q(p)}^{-2} \kappabar) = N_f \kappabar^2/(\epsilon_{p}^2 g_{q(p)}^2)$. We must require that this quantity be small in order to isolate the Yang--Mills--Higgs theory. 

In sum, the conditions for having the $\mathfrak{su}(N_f)$ sector of the probe theory described by a (tree-level) Yang--Mills--Higgs theory are:
\begin{equation}\label{YMconds}
(N_cg_p^2)^{-\frac{1}{5-p}}\ll 1~,\qquad \frac{1}{N_c} (N_c g_p^2)^{\frac{10-(p+q)}{2(5-p)}}\ll 1~, \qquad \textrm{and}\qquad \frac{N_f}{N_c} \left( N_c g_{p}^2 \right)^{\frac{q-p}{2(5-p)}} \ll 1~,
\end{equation}
assuming that the conditions for validity of the supergravity background \eqref{sugraconds}, are met. 
 Note that the region of small $v |_{z_0} =\sqrt{r^2+z_0^2}$, in which the supergravity background is no longer valid, can be avoided on the probe branes by choosing sufficiently large $z_0$. 

For any $p \leq 4$ and finite $N_f$, \eqref{YMconds} can be satisfied by taking $g_{p}^2 N_c$ and $N_c$ sufficiently large.  From the last condition we see that only when $q = p$ do we recover the ``standard'' probe limit, $N_f/N_c \ll 1$.  When $q > p$ it must be strengthened and when $q < p$ it can be relaxed -- in fact, it is possible to have $N_f$ of the same order as $N_c$ in those cases. This is one of our main results.  It demonstrates that the Yang--Mills--Higgs decoupling limit found in \cite{Domokos:2017wlb} for the D3/D5 system extends to other D$p$/D$q$ systems.

Note that the above scaling arguments also hold for the finite temperature black $p$-brane backgrounds of \cite{Horowitz:1991cd, Itzhaki:1998dd}, as the metric there contains the same overall factors as for the extremal case. The only change would be to the form of the rescaled induced metric.

\section{Classical Yang--Mills--Higgs Vacua}\label{sec:vacua}

In this section we describe some applications of the above limit, which isolates classical $(q+1)$-dimensional super-Yang--Mills--Higgs theory on the near-horizon geometry of D$p$-branes.  

In \cite{Domokos:2017wlb} we uncovered a rich class of vacua in the probe D5-brane theory arising from the D3/D5 intersection.  We review some of those results in the next subsection and generalize to other D$p$/D$(p+2)$ systems intersecting on $d=p-1$ directions. We then describe similar vacuum structures in D$p$/D$(p+4)$ systems with $d = p$ and D$p$/D$p$ systems with $d = p-2$. Some aspects of the D$p$/D$(p+4)$ analysis were previously worked out in \cite{Arean:2007nh}.
 
Supersymmetry of the probe action mandates that the Hamiltonian of the Yang--Mills--Higgs theory be a sum of squares. A key result of \cite{Domokos:2017wlb} is that the Chern-Simons action contributes terms that, together with the Yang--Mills--Higgs action, conspire so that these squares are nontrivial. Setting these particular groupings of terms to zero leads to an interesting class of field configurations that is part of the space of vacua. We will see that this phenomenon occurs in all of the examples below. 
These zero-energy vacua break no additional supersymmetry,  and have interpretations as additional color branes dissolved into the flavor branes or stretched between them.

As the nonabelian Chern-Simons action plays a key role below, we review it briefly now.  Restricting to the closed string background fields present in \eqref{NearHorizon}, the CS action takes the form \cite{Myers:1999ps}:
\begin{align}
S_{\rm CS} =&~ T_{{\rm D}q} \int \STr \bigg\{ P \left[ e^{-i(2\pi\alpha')^{-1} {\rm i}_X {\rm i}_X} C \right] \wedge e^{-i (2\pi\alpha') F} \bigg\} ~, \raisetag{24pt} \label{MyersCS}
\end{align}
where $C$ is the formal sum over even (Type IIB) or odd (Type IIA) Ramond-Ramond potentials.  The symbol ${\rm i}_X$ denotes the interior product with respect to $X^m$.  This is an anti-derivation on forms, reducing the degree by one.  Since the $X^m$ are non-commuting one has, for example, 
\begin{equation}
( {\rm i}_{X}^2 C^{(k+2)})_{M_1 \cdots M_{k}} = \half [X^m,X^n] C_{nm M_1 \cdots M_{k}}^{(k+2)}~.
\end{equation}
Different pieces in this action, when evaluated on the nontrivial D$p$-brane background, will contribute terms to $V_{2,0}^{\rm CS}$ from \eqref{V20} that complete squares in the Hamiltonian as show in detail below.

\subsection{Vacua of D$p$/D$(p+2)$}

Let's begin by reviewing some results for the D3/D5 system.  The near-horizon metric is precisely $AdS_5 \times S^5$ without any conformal factor and the $v$-dependent conformal factors in the probe brane action vanish when $p=3$. In the limit described above, $V_{2,0}$ in \eqref{DBIexp}, yields a super--Yang--Mills--Higgs theory on $AdS_4\times S^2$ (when $z_0=0$, or a deformation of it when $z_0 \neq 0$).  The four transverse scalars $\Phi^m$ are divided into a triplet $\Phi^{z_i}$, representing fluctuations of the D5-branes in the $z_i$ directions transverse to both stacks, and $\Phi^y$, representing fluctuations along the D3's. 

The classical Hamiltonian has the form
\begin{equation}
H_{\rm ym} = \frac{1}{g_{5(3)}^2}\int_{\Sigma_t} d^5 x \sqrt{-\det (g_{ab})} \left( \KK + \VV \right)
\end{equation}
where $\Sigma_T$ is a constant time slice of the asymptotically $AdS_4\times S^2$ worldvolume, and the kinetic and potential energy densities are
\begin{align}
\KK :=& - \frac{1}{2} g^{00} \Tr \left\{ F^{p}_{\phantom{p}0} F_{p0} + F^{r_i}_{\phantom{r_i}0} F_{r_i 0} + \Gbar_{mn} D_0 \Phi^m D_0 \Phi^n \right\}~, \label{kinenergy} \\
\VV :=& \Tr \bigg\{ \frac{1}{4} F^{r_i r_j} F_{r_i r_j} + \half (F^{x_1 x_2} F_{x_1 x_2} + F^{x_i r_i} F_{x_i r_i}) +  \half \Gbar_{mn} (D_{x_i} \Phi^m D^{x_i} \Phi^n + D_{r_i} \Phi^m D^{r_i} \Phi^n)  + \qquad \cr
&~ \qquad + \frac{1}{4} \Gbar_{mn} \Gbar_{m'n'} [\Phi^m, \Phi^{m'}] [\Phi^{n}, \Phi^{n'}] - \half \mu v|_{z_0}  \,\epsilon^{r_i r_j r_k} (D_{r_i} \Phi^y) F_{r_j r_k}  \bigg\}~. \label{potenergy} \raisetag{20pt}
\end{align}
Note that the final term in $\VV$, which comes from the Chern-Simons part of the action, is of the form $P(C_4)\wedge F$, where $P(C_4)$ is the nonabelian pull-back of the background $C_4$ flux to the worldvolume. Here $\epsilon^{r_ir_jr_k}$ is the covariant Levi-Civita tensor with respect to $g_{r_ir_j}$. 

The final term in \eqref{potenergy} combines with pieces of the $F^2$ and $(D\Phi^y)^2$ terms to form a complete square: 
\begin{align}\label{D3D5monopoleVacua}
\VV \supset \frac{1}{2}  \Tr \left\{ \left| F_{r_i r_j} - \mu v|_{z_0} \, \epsilon_{r_i r_j r_k} D^{r_k} \Phi^y \right|^2 \right\}~,
\end{align}
where $|\alpha_{a_1 \cdots a_p}|^2 \equiv \frac{1}{p!} \alpha_{a_1 \cdots a_p} \alpha^{a_1 \cdots a_p}$.  The conditions for zero energy thus include the Bogomolny equation for monopoles on a three-space parameterized by $\vec{r}$, with metric
\begin{equation}
g_{r_ir_j}dr^i dr^j = \frac{1}{\mu^2 (r^2+z_0^2)} d\vec{r}\cdot d\vec{r} =: \frac{1}{\mu^2 (r^2+z_0^2)} \tilde{g}_{r_i r_j} dr^i dr^j~.
\end{equation}
The unusual factor in front of the $D\Phi^y$ term in \eqref{D3D5monopoleVacua} is precisely the right one to guarantee that the non-standard Bogomolny equation with respect to $g_{r_i r_j}$ is equivalent to the standard Bogomolny equation with respect to $\tilde{g}_{r_ir_j}$, which in this case is just the flat Euclidean metric:
\begin{equation}\label{Bogoeq}
F_{r_i r_j} -  \tilde{\epsilon}_{r_i r_j r_k} D^{r_k}\Phi^y = 0~.
\end{equation}
The vaccum configurations  defined by these equations are translationally invariant with respect to the $x^\mu$ directions parameterizing the defect in the dual CFT. In particular, it would cost infinite energy to move from one point in the moduli space of these vacua to another.  

In \cite{Domokos:2017wlb} we describe the brane picture associated with these vacua.  They correspond to some of the D3-branes breaking into segments stretched between the D5-branes, which then slide outward along the $r_i$ directions while remaining parallel to the original color branes.    

We now generalize the above analysis to D$p$/D$(p+2)$ intersections, in which the background D$p$'s extend along one direction transverse to the flavor D$(p+2)$ branes. As above, the Chern-Simons action contributes a cross term between the transverse $\Phi^y$ scalar and $F_{r_i r_j}$:
\begin{align}
S_{\rm CS} \supset&~ T_{{\rm D}(p+2)} \int \Tr \left\{ P[C^{(p+1)}] \wedge (2\pi\alpha') (-iF)\right\}\nonumber\\
=&  -T_{{\rm D}(p+2)} \epsilon_p (2\pi\alpha')  \int d^{p+3}x  \frac{1}{2} C_{01\cdots (p-1) y} \Tr\left\{ D_{r_i} \Phi^y F_{r_j r_k}\right\}  \varepsilon^{01\cdots (p-1) r_i r_j r_k} \nonumber\\
=& \frac{\epsilon_p T_{{\rm D}(p+2)} \lambda_C (2\pi\alpha')}{2} \int d^{p+3}x \sqrt{-\det (g_{ab})} (\mu v|_{z_0})^{\frac{(p-4)(p-7)}{2(5-p)}} \times \cr
&~ \qquad \qquad \qquad \qquad \qquad \qquad \times  \epsilon^{r_i r_j r_k} \Tr \left\{ D_{r_i}\Phi^y F_{r_jr_k}\right\}~.
\end{align}
In the second line, $\varepsilon^{01\cdots (p-1) r_1 r_2 r_3}=1$ is the Levi-Civita tensor density, while $\epsilon^{r_ir_jr_k}$ in the third line is the covariant Levi-Civita tensor with respect to the metric $g_{r_ir_j}$.  $\lambda_C$ is defined in \eqref{NearHorizon}, while the factor of $\epsilon_p$ in the second line comes from converting the fluctuation scalar $X^y$ to the mass-dimension-one version, $\Phi^y$. 

Combining the Chern-Simons contribution with the rest of the Yang--Mills--Higgs term \eqref{V20}, and making use of \eqref{epsp} and the relation
\begin{equation}\label{epsrel}
\lambda_C = \lambda_G^{(p+1)/2} \lambda_\phi^{-1} ~,
\end{equation}
we find that the Chern-Simons term allows us to complete a square in the Hamiltonian:
\begin{align}
H\supset&~ \frac{1}{g_{p+2(p)}^2} \int d^{p+3}x \sqrt{-\det(g_{ab})}(\mu v|_{z_0})^{\frac{(p-3)(p-7)}{2(5-p)}} \times \nonumber\\
&\qquad\qquad\qquad \times\frac{1}{2}\Tr\left\{ \left| F_{r_ir_j}- (\mu v|_{z_0})^{\frac{(7-p)}{2(5-p)}} \epsilon_{r_ir_jr_k} D^{r_k}\Phi^y\right|^2 \right\}~.
\end{align}
Performing an additional rescaling of the induced metric, we again find the Bogomolny equations of \eqref{Bogoeq}, except that now $\tilde{g}_{r_ir_j}$ is given by
\begin{equation}\label{monometric}
\tilde{g}_{r_ir_j}dr^i dr^j = (\mu v|_{z_0})^{\frac{(p-3)}{2(5-p)}} \left[ \left( \frac{r^2+\alpha_p^{-2}z_0^2}{r^2+z_0^2}\right) dr^2+ \frac{r^2}{\alpha_{p}^2}d\Omega_2^2\right] ~.
\end{equation}
This reduces to the flat Euclidean metric for $p=3$.  When $z_0=0$, \eqref{monometric} is conformal to a cone over $S^2$.  For $p<3$, the cone has an excess angle, while for $p=4$ it is $\mathbb{R}^3/\mathbb{Z}_2$.  If $z_0\ne 0$, the space is conformally asymptotic to these cones, but is capped off smoothly at $r=0$.  The interpretation of these vacua from the intersecting brane picture is as we described above:  a number of D$p$-branes parallel to the original color branes break into segments stretched between the D$(p+2)$'s.

\subsection{Vacua of D$p$/D$(p+4)$}\label{sec:instantons}

Next, we turn to vacua of the probe D$(p+4)$-brane theory in the near-horizon background of D$p$-branes, where a similar story plays out.  
In this class of examples the flavor branes span all directions parallel to the color branes, so there are no $y$-type coordinates.  

The Chern-Simons action for the probe branes contains a term of the form $P(C^{(p+1)})\wedge \Tr (F\wedge F)$:
\begin{align}\label{CSppplus4}
S_{{\rm CS}} &\supset - T_{{\rm D}(p+4)} \int C^{(p+1)}\wedge \left( \frac{(2\pi\alpha')^2}{2} \Tr (F\wedge F )\right) \nonumber\\
& =  T_{{\rm D}(p+4)} (2\pi\alpha')^2\lambda_C \int d^{p+5}x \sqrt{-\det (g_{ab})} \, (\mu v |_{z_0})^{\frac{(p-3)(p-7)}{2(5-p)}} \Tr \left\{ \frac{1}{8}\epsilon^{r_i r_j r_k r_l} F_{r_i r_j} F_{r_k r_l}\right\}~,
\end{align}
where $\epsilon^{r_i r_j r_k r_l} $ is the Levi-Civita tensor with respect to the rescaled induced metric $g_{ab}$ restricted to the $\vec{r}$ part of the worldvolume. 

Combining this term with the gauge kinetic terms from \eqref{DBI}, and using \eqref{epsrel},
we find that the Hamiltonian
\begin{align}
H&\supset \frac{1}{g_{p+4(p)}^2}\int_{\Sigma_t} d^{p+4}x \sqrt{-\det (g_{ab})} (\mu v |_{z_0})^{\frac{(p-3)(p-7)}{2(5-p)}} \times \cr
&~ \qquad \qquad \qquad \qquad \times \Tr \left\{  \frac{1}{4} F_{r_i r_j} F^{r_i r_j} -\frac{1}{8} \epsilon^{r_i r_j r_k r_l} F_{r_i r_j} F_{r_k r_l} \right\} \nonumber \\
&= \frac{1}{4g_{p+4(p)}^2}\int_{\Sigma_t} d^{p+4}x \sqrt{-\det (g_{ab})} (\mu v |_{z_0})^{\frac{(p-3)(p-7)}{2(5-p)}} \Tr\left\{ \left| F_{r_i r_j}  -\frac{1}{2} \epsilon_{r_i r_j r_k r_l} F^{r_k r_l} \right|^2 \right\}~. \qquad \quad
\end{align}
Setting the Hamiltonian to zero leads to the equation for instantons on the space parameterized by $\vec{r}$, with metric
\begin{align}\label{instmetric}
g_{r_i r_j}d{r^i}d{r^j} = (\mu v |_{z_0})^{\frac{(p-3)}{(5-p)} -2} \left[\left( \frac{r^2 + \alpha_{p}^{-2} z_{0}^2}{r^2 + z_{0}^2} \right) dr^2 +  \frac{r^2}{\alpha_{p}^{2}}d\Omega_{3}^2\right]~.
\end{align}
The instanton equation is conformally invariant, so this is equivalent to instantons on the space
\begin{equation}\label{4cone}
\tilde{g}_{r_i r_j} dr_i dr_j = \left(\frac{r^2 + \alpha_{p}^{-2} z_{0}^2}{r^2 + z_{0}^2} \right) dr^2 +  \frac{r^2}{\alpha_{p}^{2}}d\Omega_{3}^2~,
\end{equation}
which is asymptotic to a cone over $S^3$.  Hence the space of classical vacua of the D$(p+4)$ probes in the near-horizon geometry of the D$p$-branes includes moduli spaces of instantons on \eqref{4cone}.

Let us put this discussion in context.  It was an important result from the early days of brane dynamics that D$p$-branes coincident with D$(p+4)$-branes appear as instantons in the worldvolume theory of the D$(p+4)$-branes \cite{Witten:1995gx,Douglas:1995bn}.  Starting with the brane system in flat space (and decoupling gravity), one can isolate the instanton degrees of freedom from either point of view -- that of the D$(p+4)$ worldvolume theory or the D$p$ worldvolume theory.  The former is achieved via a supersymmetric instanton collective coordinate ansatz.  In the latter case, the low energy dynamics of the $p$-$p$ and $p$-$(p+4)$ strings furnish a gauged linear sigma model.  The D- and F-flatness conditions for supersymmetric vacua implement the ADHM construction \cite{Atiyah:1978ri} of the instanton \cite{Witten:1994tz,Douglas:1996uz,Barbon:1997ak,Dorey:1999pd}.  Integrating out the ADHM variables results in the same nonlinear sigma model obtained from the collective coordinate construction.  (See the review, \cite{Dorey:2002ik}, for a nice summary of these results.)  

Note that, in the old story just described, the instanton moduli space plays very different roles in each of these two points of view.  From the point of view of the (flat space) D$(p+4)$-brane theory the moduli space parameterizes a set of half BPS codimension-four defect solutions carrying \emph{nonzero} energy density.  From the point of  view of the D$p$-brane theory the moduli space parameterizes a set of zero-energy vacua.  There is no contradiction here because the low-energy limits that decouple gravity and reduce the worldvolume theories to their respective gauge theories are different for the D$(p+4)$-brane and the D$p$-brane.  One of the two Yang--Mills couplings can be held fixed in the limit, but not both.

What we have shown in this paper is that holography provides a \emph{third} perspective on the instanton moduli space.  In the low-energy limit that holds the D$p$-brane Yang--Mills coupling fixed, there is a dual description of that theory in terms of a string theory with probe D$(p+4)$-branes in the near horizon geometry of the D$p$-branes.  The same moduli space of vacua that we discussed above must exist in that theory as well.  The description of the dual string theory we've developed here, in terms of a Yang--Mills--Higgs theory on the probe branes weakly coupled to supergravity, only exists in the regime of parameter space specified by \eqref{YMconds}.  In particular this implies that the number of D$p$-branes in the original setup, $N_c$, must be large.  Hence the space of vacua we observe in the probe brane theory -- the instantons on the asymptotically conical space \eqref{4cone} -- must be a sublocus of the total space of vacua of the D$p$-brane worldvolume theory.  

It would be interesting to see how the description of this sublocus emerges in the holographically dual defect QFT, where the ADHM construction generates this moduli space of vacua.  How will the background geometry these instantons live on emerge?  We anticipate that it comes about through a nontrivial saddle point approximation of the ADHM-like equations that couple the massless $p$-$p$ and $p$-$(p+4)$ string degrees of freedom.  This would be somewhat analogous to how the $AdS_5\times S^5$ geometry emerges from instanton moduli space at large $N_c$, as explained in \cite{Dorey:1999pd}.  Here, however, we would be considering large instanton charge and finite rank gauge group, since the instantons are the color branes rather than the flavor branes.  We comment a bit further on this in the conclusion.

\subsection{Vacua of D$p$/D$p$}

Having found systems in which the vacuum structure includes monopole moduli spaces (D$p$/D$(p+2)$) and instantons (D$p$/D$(p+4)$), we now consider the D$p$/D$p$ system, in which the vacuum structure includes vortices.  The D$p$ flavor branes are extended along two directions in which the D$p$ color brane are not and viceversa, so $\vec{r}=(r_1,r_2)$ and $\vec{y}=(y_1, y_2)$. The number of shared spatial directions $d=p-2$ is arbitrary. Again we find that the Chern-Simons action adds the cross terms that complete squares in the Yang-Mills action. This time, however, there is a square involving $F_{r_ir_j}$ and the commutator $[\Phi^{y_i},\Phi^{y_j}]$, and, separately, squares involving the covariant derivatives $D_{r_i}\Phi^{y_i}$.  

The relevant terms from the CS action are 
\begin{align}
S_{\rm CS} \supset &~  T_{Dp}\int \Tr \left\{ P[C^{(p+1)}] - P \left[ \mathrm{i}_X^2 C^{(p+1)}\right] \wedge F\right\} \cr
=&~ T_{Dp} \epsilon_p^2 \lambda_C \int d^{p+1}x \sqrt{-\det(g_{ab})} \, (\mu v |_{z_0})^{\frac{(p-3)(p-7)}{2(5-p)}} \times \cr
&~  \qquad \qquad \quad  \times \epsilon^{r_i r_j} \epsbar_{y_i y_j} \Tr  \left\{ \half D_{r_i} \Phi^{y_i} D_{r_j}\Phi^{y_j}+ \frac{1}{4} [\Phi^{y_i},\Phi^{y_j}] F_{r_ir_j} \right\} 
\end{align}
where $\epsilon_{r_ir_j}$ is the Levi--Civita tensor with respect to the rescaled induced metric $g_{r_ir_j}$ and $\epsbar_{y_i y_j}$ is the Levi-Civita tensor with respect to the rescaled metric $\Gbar_{y_i y_j}$. When added to the DBI action, this yields a Hamiltonian containing the following set of terms:
\begin{align}\label{vortexH}
H \supset&~ \frac{1}{g_{p(p)}^2} \int_{\Sigma_t} d^px \sqrt{-\det (g_{ab})} \, (\mu v |_{z_0})^{\frac{(p-3)(p-7)}{2(5-p)}} \times \cr
&~ \qquad \times \Tr \bigg\{ \frac{1}{4} F_{r_i r_j} F^{r_i r_j} - \half  \epsilon^{r_i r_j} \epsbar_{y_i y_j} F_{r_i r_j} [\Phi^{y_i},\Phi^{y_j}]  + \frac{1}{4} \Gbar_{y_i y_j} \Gbar_{y_k y_l} [\Phi^{y_i}, \Phi^{y_k}] [\Phi^{y_j},\Phi^{y_l}] + \cr
&~ \qquad \qquad \quad + \frac{1}{2} \left( \Gbar_{y_i y_j} D_{r_i} \Phi^{y_i} D^{r_i} \Phi^{y_j} - \epsilon^{r_i r_j} \epsbar_{y_i y_j} D_{r_i} \Phi^{y_i} D_{r_j} \Phi^{y_j}  \right) \bigg\}~, \qquad 
\end{align}
where we have made use of \eqref{epsrel} and \eqref{gqpdef}.

It is easy to see that the first three terms in \eqref{vortexH} form a complete square.  The remaining two terms can also be written as a sum of squares; in fact there is an $\mathrm{SO}(2)$ family of possibilities as we now show. Let $\ebar^{\underline{y_i}}_{\phantom{y_i}y_i}$ denote the components of an orthonormal co-frame in the $y$-directions such that $\ebar^{\underline{y_i}}_{\phantom{y_i}y_i} \ebar^{\underline{y_j}}_{\phantom{y_j}y_j} \delta_{\underline{y_i y_j}} = \Gbar_{y_i y_j}$, and similarly let $e_{\underline{r_i}}^{\phantom{r_i}r_i}$ denote the components of an orthonormal frame in the $r$-directions such that $e_{\underline{r_i}}^{\phantom{r_i}r_i} e_{\underline{r_j}}^{\phantom{r_j}r_j} \delta^{\underline{r_i r_j}} = g^{r_i r_j}$.  Let $\RR_{\underline{y_i}}^{\phantom{y_i}\underline{r_i}}$ be an orthogonal transformation with determinant minus one.  Then, thanks to the identity
\begin{align}
& \ebar^{\underline{y_i}}_{\phantom{y_i}y_i} \ebar^{\underline{y_j}}_{\phantom{y_j}y_j} \RR_{\underline{y_i}}^{\phantom{y_i}\underline{r_i}} \RR_{\underline{y_i}}^{\phantom{y_i}\underline{r_i}} \left( e_{\underline{r_i}}^{\phantom{r_i}r_i} e_{\underline{r_j}}^{\phantom{r_j}r_j} + e^{\underline{r_k} r_i} \varepsilon_{\underline{r_k r_i}} e^{\underline{r_l} r_j} \varepsilon_{\underline{r_l r_j}} \right) = \qquad \qquad \qquad \qquad \qquad \qquad \cr
& \qquad \qquad \qquad \qquad \qquad \qquad = \ebar^{\underline{y_i}}_{\phantom{y_i}y_i} \ebar^{\underline{y_j}}_{\phantom{y_j}y_j} \RR_{\underline{y_i}}^{\phantom{y_i}\underline{r_i}} \RR_{\underline{y_i}}^{\phantom{y_i}\underline{r_i}} \left( \delta_{\underline{r_i r_j}} g^{r_i r_j} + \varepsilon_{\underline{r_i r_j}} \epsilon^{r_i r_j} \right) \cr 
& \qquad \qquad \qquad \qquad \qquad \qquad = \ebar^{\underline{y_i}}_{\phantom{y_i}y_i} \ebar^{\underline{y_j}}_{\phantom{y_j}y_j}  \left( (\RR \RR^T)_{\underline{y_i y_j}} g^{r_i r_j} + (\RR \varepsilon \RR^T)_{\underline{y_i y_j}} \epsilon^{r_i r_j}  \right) \cr
& \qquad \qquad \qquad \qquad \qquad \qquad = \Gbar_{y_i y_j} g^{r_i r_j} - \epsbar_{y_i y_j} \epsilon^{r_i r_j} ~,
\end{align}
one can rewrite \eqref{vortexH} in the form
\begin{align}\label{vortexH2}
H \supset&~ \frac{1}{g_{p(p)}^2} \int_{\Sigma_t} d^px \sqrt{-\det (g_{ab})} \, (\mu v |_{z_0})^{\frac{(p-3)(p-7)}{2(5-p)}} \times \cr
&~ \qquad  \times \Tr \bigg\{ \frac{1}{2}  \left[ \left( \ebar^{\underline{y_i}}_{\phantom{y_i}y_i} \RR_{\underline{y_i}}^{\phantom{y_i}\underline{r_i}} e_{\underline{r_i}}^{\phantom{r_i} r_i} D_{r_i} \Phi^{y_i} \right)^2 + \left( \ebar^{\underline{y_i}}_{\phantom{y_i}y_i} \RR_{\underline{y_i}}^{\phantom{y_i}\underline{r_i}} \varepsilon_{\underline{r_i r_j}} e^{\underline{r_j}r_i} D_{r_i} \Phi^{y_i} \right)^2  \right]  + \cr 
&~ \qquad \qquad \quad + \frac{1}{8} \left( \epsbar^{y_i y_j} F_{r_i r_j} -\epsilon_{r_i r_j} [\Phi^{y_i}, \Phi^{y_j}] \right)^2 \bigg\}~, \qquad 
\end{align}
where on the last term all free indices are contracted with the appropriate metric after squaring.  Thus field configurations satisfying the first order equations
\begin{align}\label{vortexeqns}
0 =&~ \epsbar^{y_i y_j} F_{r_i r_j} - \epsilon_{r_i r_j} [\Phi^{y_i}, \Phi^{y_j}]~, \cr
0 =&~  \ebar^{\underline{y_i}}_{\phantom{y_i}y_i} \RR_{\underline{y_i}}^{\phantom{y_i}\underline{r_i}} e_{\underline{r_i}}^{\phantom{r_i} r_i} D_{r_i} \Phi^{y_i} ~, \cr
0 =&~  \ebar^{\underline{y_i}}_{\phantom{y_i}y_i} \RR_{\underline{y_i}}^{\phantom{y_i}\underline{r_i}} \varepsilon_{\underline{r_i r_j}} e^{\underline{r_j}r_i} D_{r_i} \Phi^{y_i}  ~,
\end{align}
yield zero-energy vacua.

The choice of $\RR \in \mathrm{O}(2)_-$ has no consequence for the space of vacua.  We can always orient the axes in $y$-space to arrange that, \eg, $\RR = \diag(1,-1)$.  However, positive-energy BPS configurations above these vacua might depend on the asymptotic values of $\Phi^{y_i}$ as $r \to \infty$, leading to different BPS bounds depending on the choice of $\RR$.  Then one would need to vary $\RR$ to achieve the strongest bound.

When $p=3$ the system \eqref{vortexeqns} is equivalent to Hitchin's equations \cite{MR887284} on $\mathbbm{R}^2$.  Taking the $y$-space co-frame and the $r$-space frame to be $\ebar^{\underline{y_i}}_{\phantom{y_i}y_i} = (\mu v) \delta^{\underline{y_i}}_{\phantom{y_i} y_i}$ and $e_{\underline{r_i}}^{\phantom{r_i}r_i} = (\mu v) \delta_{\underline{r_i}}^{\phantom{r_i} r_i}$, and setting
\begin{equation}
\Phi^i := (\RR^T)^{\underline{r_i}}_{\phantom{r_i}\underline{y_j}} \delta^{\underline{y_j}}_{\phantom{y_j} y_j} \Phi^{y_j}~, \qquad i = 1,2~,
\end{equation}
one finds
\begin{align}
0 =&~ F_{r_1 r_2} + [\Phi^1,\Phi^2]~, \cr
0 =&~ D_{r_1} \Phi^1 + D_{r_2} \Phi^2~,  \qquad \qquad \textrm{(for $p=3$)}  \cr
0 =&~ D_{r_1} \Phi^2 - D_{r_2} \Phi^1~.
\end{align}
Setting $\Phi = \Phi^1 - i \Phi^2$, the first equation takes the form $F_{r_1 r_2} = \frac{i}{2} [\Phi, \Phi^\ast]$ while the second two equations are equivalent to $(D_{r_1} + i D_{r_2})\Phi = 0$ and its conjugate.  To write these equations in the usual form of Hitchin's equations, introduce a complex coordinate $z = r_1 + i r_2$ and a one-form $\phi = \half \Phi dz$.  Then we have
\begin{equation}
F = - [\phi, \phi^\ast]~, \qquad \Dbar \phi = 0~,
\end{equation}
where $\Dbar = d\zbar D_{\zbar}$ with $D_{\zbar} = \half (D_{r_1} + i D_{r_2})$.

Solutions to this system on $\mathbbm{R}^2$ have been studied starting in \cite{Lohe:1977ht,Saclioglu:1980ux,Saclioglu:1984vh}, where it was shown that solutions with finite Higgs fields at infinity must necessarily have singularities.  Such solutions provide limiting configurations for studying the asymptotic regions of the moduli space of solutions to Hitchin's equations on a compact surface \cite{Gaiotto:2009hg,MR3544281}. Alternatively, one can obtain smooth solutions in the plane if one allows the Higgs fields to diverge polynomially in $z$ \cite{MR3465983}.  

The brane picture provides some intuition for these observations.  These solutions are describing D3-branes ending on a system of D3-branes in co-dimension two.  In such low co-dimension, the pull of D3-brane segments on the D3-branes they are stretched between will cause  runaway behavior at infinity, unless it is balanced by an external pull on the stack.  The external pull would be due to semi-infinite D3-branes and correspond to the singularities in the fields.  It would be interesting to see if the brane picture can provide a detailed accounting of the moduli of singular solutions to Hitchin's equations on $\mathbbm{R}^2$, as has been done for singular monopoles on $\mathbbm{R}^3$ in \cite{Moore:2014gua}.      

For $p \neq 3$ the system \eqref{vortexeqns} is distinct from Hitchin's equations.  In order to arrive at the latter from BPS conditions in super--Yang--Mills--Higgs theory on curved space, one generally needs to perform a topological twist of the theory (see \eg\ \cite{Bershadsky:1995vm,Kapustin:2006pk}); this is not the scenario we are studying here.

\section{Summary and Directions for Future Work}\label{sec:concl}

In this paper, we derived the conditions \eqref{YMconds} necessary for isolating a classical supersymmetric Yang--Mills--Higgs theory on $N_f$ probe D$q$-branes in the near-horizon geomtery of $N_c$ D$p$-branes.  We focused on D$p$ backgrounds possessing local field theory duals ($p\le 4$), with probe configurations that lend themselves to holography (probe branes that extend all the way to the AdS boundary). We found that by tuning $N_c$, $N_f$, and the dimensionless 't Hooft coupling $N_c g_p^2$, it is always possible to find a corner of parameter space in which the theory can be approximated by classical Yang--Mills--Higgs on a rigid background. 

We also noted that supersymmetry permits a one-parameter family of deformations of the AdS background -- equivalent to separating the D$p$ and D$q$ stacks along a direction transverse to both. This is dual to a relevant deformation in the dual field theory. While we focused here on supersymmetric intersections, the analysis of Section \ref{sec:YM} should generalize straightforwardly to non-supersymmetric cases with holographic interpretations, like finite-temperature D$p$ backgrounds, or ones in which the D$p$'s wrap a compact cycle. In each of these cases, the modification to the background introduces an additional scale (the temperature for the former, and a  Kaluza-Klein scale for the latter). These scales should not change how we take the near-horizon limit, however. 

Another result of the current work is that, in the limit \eqref{YMconds} the spaces of zero-energy vacua of these gauge theories include moduli spaces of monopoles, instantons, and vortices on curved backgrounds. In Section \ref{sec:vacua}, we exhibited the first-order equations these must satisfy.  

It would be interesting to uncover these vacua in the holographically dual field theories for several reasons.  

First, even though the gauge theory on the probe branes decouples from the closed string degrees of freedom in a modified version of the probe limit,  the {\em vacuum} structure of the probe brane theory relies on cooperation with the background, which provides not only the curved metric, but the Ramond-Ramond terms that give rise to the Chern-Simons contribution at quadratic order in the action. It is only the fluctuations of the bulk modes that decouple from fluctuations of open string modes around these vacua. Similarly, in the dual field theory one would expect that the defect interaction with the ambient field theory generates a class of vacua, with decoupling of fluctuations only.

Second, our results, together with the standard description of lower-dimensional branes as solitons in higher-dimensional brane worldvolumes, point towards an intriguing new regime in which to study soliton moduli spaces.  We mentioned this in the context of the instanton vacua in subsection \ref{sec:instantons}, but the following remarks apply more generally.

We found that the spaces of vacua of the Yang--Mills--Higgs theories on the probe branes include moduli spaces of solitons on curved backgrounds.  The brane picture indicates that these moduli spaces might capture a sublocus of the ordinary (\ie\ flat-space) soliton moduli spaces at large charge (large $N_c$).  The reason we cannot be certain of this is that the path connecting the two pictures -- solitons on a curved spaced versus a sublocus of ordinary many-soliton moduli space -- goes through the configuration space of string theory and includes replacing an $O(1)$ fraction of the ordinary solitons by the geometry they generate.

Studying the vacua of these theories from the field theory side of the holographic correspondence could shed light on this question.  In that description, the vacua should arise from the ADHM or Nahm-like construction of solitons, just as they do in the standard intersecting brane picture -- see \eg\ \cite{Diaconescu:1996rk, Kapustin:1998pb}.  We expect that the curved geometry on which the solitons live should arise from a saddle point analysis of the Nahm-like equations.  Indeed an analogous result was beautifully demonstrated for the case of D-instantons in the standard correspondence between $\NN = 4$ super-Yang--Mills and string theory on $AdS_5 \times S_5$ \cite{Banks:1998nr,Dorey:1999pd}.  The difference here is that the probe branes and the background-generating branes switch roles, so that one would be considering a saddle point approximation at large soliton charge rather than large gauge group rank.  We intend to investigate this possibility in the future.

{\bf Acknowledgements:}  This work was performed in part at the Aspen Center for Physics, which is supported by National Science Foundation grant PHY-1607611. SKD and ABR also thank Ibrahima Bah and Daniel Robbins for helpful discussions.



\providecommand{\href}[2]{#2}\begingroup\raggedright\endgroup

\end{document}